\documentstyle[preprint,prb,eqsecnum,aps,epsf]{revtex}
\begin{document}
\tighten
\draft
\preprint{IUCM96-001}

\title{\centerline{Spontaneous coherence and the quantum Hall Effect}
       \centerline{in triple-layer electron systems}}

\author{C.~B.~Hanna and A.~H.~MacDonald}
\address{Department of Physics, Indiana University,
Bloomington, Indiana~~47405\\}

\date{\today}

\maketitle

\begin{abstract}

We investigate spontaneous interlayer phase coherence and the 
occurrence of the quantum Hall effect in triple-layer
electron systems.  Our work is based on a simple
tight-binding model that greatly facilitates calculations
and whose accuracy is verified by comparison with recent 
experiments.  By calculating the ground state 
in an unrestricted Hartree-Fock approximation and 
the collective-mode spectrum in a time-dependent
Hartree-Fock approximation, 
we construct a phase diagram delimiting regions
in the parameter space of the model where the integer quantum Hall
effect occurs in the absence of interlayer tunneling.

\end{abstract}

\pacs{PACS numbers: 73.20.Dx, 75.10.-b, 64.60.Cn}

\section{Introduction}
\label{sec:intro}

Interactions among particles play an especially important role in
two-dimensional (2D) systems in the quantum Hall
(QH) regime because the
kinetic energy is quenched.  Kinetic energy eigenstates bunch into
quantized Landau levels with macroscopic degeneracy
$N_{\phi} = B_\perp A/\Phi_0$.
(Here $A$ is the cross-sectional area of the system,
$B_\perp$ is the 
magnetic field strength, and $\Phi_0 = hc/e$ is the magnetic flux
quantum.) The fractional QH effect (QHE),\cite{tsui}
which occurs when an orbitally degenerate
Landau level has a fractional filling factor
$\nu_{\rm{T}} \equiv N_e/N_{\phi}$,
where $N_e$ is the total number of electrons,
arises from strongly correlated states produced entirely
by interactions.\cite{rbl,prange}

Interactions can play an important role even at integer values 
of $\nu_{\rm{T}}$, if at low energies the system has additional 
degrees of freedom.  An important example occurs 
for an isolated 2D electron layer at $\nu_{\rm{T}} =1$,
when the Zeeman energy is so small that the spins are not
completely frozen.
It turns out in this case that the ground state is
completely spin polarized, and that the energy gap for charged
excitations (which gives the QHE) is finite,
even in the limit of vanishing Zeeman splitting.
However, there is strong experimental evidence
that for $\nu_{\rm{T}}$ close,
but not equal, to $1$, the ground state contains a large number
of flipped spins.\cite{barrett}  This property of the single-layer
system, which was anticipated
theoretically,\cite{rezayi,sondhi,fertighf}
is best understood by recognizing that
the ground state at $\nu_{\rm{T}}=1$
is a 2D ferromagnet with spontaneous spin-polarization.
2D ferromagnets
have stable finite-energy topologically charged spin-texture 
excitations, commonly known as skyrmions.\cite{rajaraman}
A unique aspect of these QH ferromagnets, first appreciated by 
Sondhi {\it et al.},\cite{sondhi} is that
their skyrmions also carry a unit electrical charge.
It is the presence of these unusual charged objects
with many reversed spins in their interior that is responsible
for the rapid decline in the spin polarization that occurs as
$\nu_{\rm{T}}$ moves away from $1$.   Many aspects of the physics are
very similar\cite{gmchap,wenzee,ezawa,yang,dlqhtheo} 
when the additional degree of freedom comes from a second 
electron layer, rather than from the two spin states available
to spin-$1/2$ particles.  The role played by the Zeeman energy
is taken over in this case by the interlayer hopping amplitude $t$,
the broken symmetry is spontaneous interlayer phase coherence 
rather than spin magnetization, and a QHE can
occur\cite{dlexpt,murphy} at $\nu_{\rm{T}}=1$, even when $t=0$. 
The combination of the integer QHE and spontaneous
broken symmetry in the ground state gives rise to charged 
order-parameter textures and other new physics.\cite{gmchap}

In this paper, we consider the case of triple-layer electron
systems (TLESs).
Our work is motivated primarily by recent progress in fabricating 
high-mobility electron systems, which has
made it possible to study these systems
experimentally.\cite{jo,lay,lay2} Triple-layer systems in
strong magnetic fields have been studied previously,\cite{macdmulti}
but the possibility of QHEs associated with 
spontaneously broken symmetries,\cite{wenzee} on which we focus here,
was not explicitly addressed.  In Sec. II of this paper, we 
introduce the tight-binding model that we use to describe TLESs
in terms of a small number of parameters.
We make some estimates of the size of the model parameters
in Sec. III, and discuss some
electrostatic considerations that are very important in
interpreting experiments.  We test our model and demonstrate 
the possibility of determining its parameters by comparing with
some recent experiments at weak magnetic fields.
In Sec. IV, we discuss unrestricted
Hartree-Fock  approximations (HFAs) for the ground state of TLESs 
in a strong magnetic field.  The HF wave functions allow for
the possibility of spontaneous interlayer phase coherence and are 
generalizations of those proposed previously\cite{moon}
for double-layer systems at $\nu_{\rm{T}} =1$:
\begin{equation}
|\Psi\rangle = \prod^{N_\phi}_X
\left ( e^{i\phi_1} {\hat c}^\dagger_{1X} +
e^{i\phi_2} {\hat c}^\dagger_{2X} \right ) |0\rangle .
\label{eq:psi2}
\end{equation}
Here $X$ is the guiding-center label
for orbital states within a Landau 
level in the Landau gauge, which we use throughout this paper.
This many-particle
wave function exhibits interlayer coherence, because all electrons
occupy states that are a linear superposition of
layer $j=1$ and layer $j=2$.
When $\phi_1=\phi_2$, Eq.~(\ref{eq:psi2}) represents a full
Landau level formed from the symmetric combination of isolated
layer states, and is evidently the exact ground state
for a system of 
noninteracting electrons, when tunneling between the 
layers is described in a tight-binding model.  
It turns out that it is also the exact ground state in the presence
of repulsive interactions, even without tunneling,
when the interactions are independent of the layer index $j$
(i.e., in the limit that the layer separation $d\rightarrow 0$).
When the interlayer Coulomb interactions are different
from the intralayer interactions ($d>0$),
Eq.~(\ref{eq:psi2}) is still
a good variational wave function because phase coherence 
guarantees good interlayer electronic correlations and
thereby lowers the interlayer Coulomb interaction
energy.\cite{gmchap}
Only when the layers are widely separated does Eq.~(\ref{eq:psi2})
become a poor variational wave function; eventually, it becomes
much more important to have good intralayer correlations than to
have good interlayer correlations.  In this case, the ground state
no longer has spontaneous interlayer correlations and
the QHE will not generally occur at $\nu_{\rm{T}}=1$
in the absence of interlayer tunneling.
(An exception occurs when the electrostatic environment is
consistent with having QHEs in which interlayer correlations play
no role, for example by having $\nu =1/3$ in each layer.)
In Sec. V,
we evaluate the collective-mode dispersion of a TLES in the 
time-dependent HF approximation (TDHFA).
In the absence of interlayer tunneling,
the energy in the double-layer
case is independent of the global phases $\phi_j$.
The $U(1)$ symmetry associated with
$\phi_{12}\equiv\phi_1-\phi_2$ is broken
in the ground state of the double-layer system,
giving rise to a gapless Goldstone mode associated
with slow spatial variations of $\phi_{12}$.\cite{wenzee,yang}
The energy cost of the spatial variations of $\phi_{12}(X)$
is due to the loss of interlayer Coulomb exchange energy.
We find that in triple-layer systems, two Goldstone modes
occur, associated with the two independent relative phases.
We use the stability of the collective modes as an indication
that the HF variational wave function is still good,
and use this criterion to map out the region in the model's
parameter space where we expect a triple-layer integer QHE to occur
with spontaneous interlayer phase coherence in the ground state.
Section VI contains some brief concluding remarks.  A 
future work will discuss charged order-parameter textures
and the effect of tilted magnetic fields in triple-layer 
systems, using a field-theoretic approach.

\section{Model for Triple-Layer Systems}
\label{sec:model}

Triple-layer QH systems have been realized experimentally
by Shayegan and co-workers.\cite{jo,lay,lay2}
In order to observe the QHE, the mobility of these samples
needs to be very high, which generally requires that they
be remotely doped.
The simplest possible theory of the triple-layer system is one
that regards it as a macroscopic metal, and therefore requires that
the volume between the right and left layers be an equipotential.
In such a theory, the Poisson equation allows no
charge in the central layer; repulsive interactions
between the electrons cause them to migrate to the left and right
layers.  Obviously, this theory is incomplete, but 
its indication that electrons tend to avoid the middle layer
is telling, and this tendency must be countered if true
triple-layer systems are to be realized.  In the TLESs
grown by Shayegan's group, the middle layer is wider 
than the outside layers and therefore has a smaller size-quantization
energy.

The Shayegan group has demonstrated\cite{jo,lay,lay2}
that at zero magnetic field,
the partitioning of electron density between
the three layers as a function of gate voltage can be accurately
rendered using a three-dimensional (3D) density-functional based 
independent-electron approximation.
In order to describe the many-body
physics of these systems, which is essential at strong magnetic
fields, we require a relatively simple model for the growth-direction
($z$-direction) spatial degree of freedom.
We proceed by generalizing the 
approach commonly used for double-layer 2D electron systems.
We assume that only the lowest electric subband is 
important in each 2D layer, and use a tight-binding
description for the $z$-direction degree of freedom,
with tunneling amplitude $t$ between neighboring 
layers, and no direct tunneling between left and right layers.
(Remote tunneling can be readily incorporated in the model, if 
the experimental situation warrants introducing this complication.) 
For the calculations reported here,
we neglect the finite width of the subband wave functions
in each layer, but these can easily be modeled if necessary in
specific systems by adding form-factor corrections\cite{afs}
to the effective electron-electron interaction.
We assume that the subband energy in the middle layer
(relative to the local electrostatic potential)
differs from the subband energies in the side layers
(also relative to the local electrostatic potential)
by $\epsilon_b$.

\begin{figure}[t]
\epsfxsize4.0in
\centerline{\epsffile{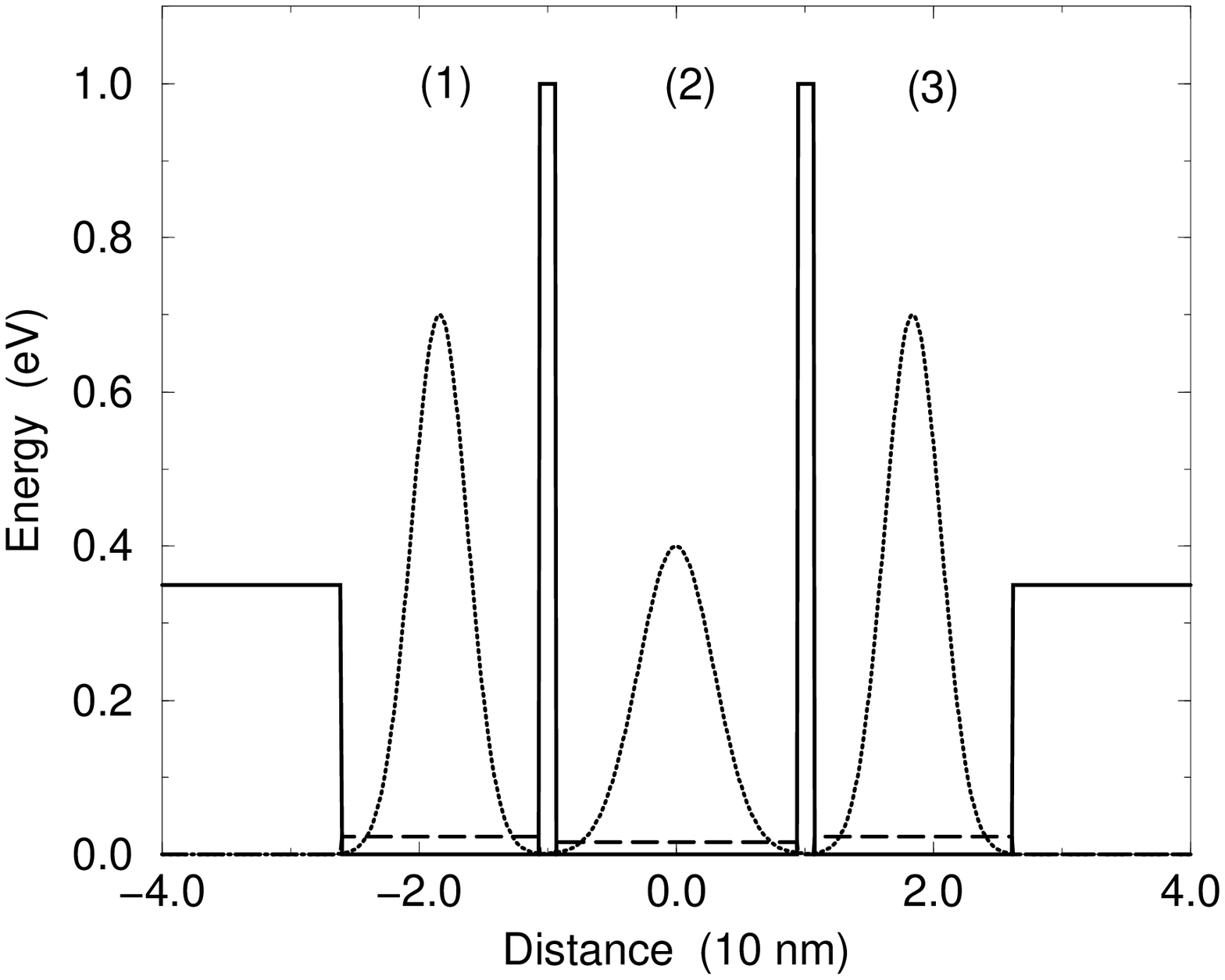}}
\caption{Schematic illustration of the triple-layer electron system
of Ref.~18.
The solid lines represent the energy of the confining barriers,
and the long-dashed lines are the energies of the lowest-energy
quantum state for a given well.
The dotted curves represent electron densities,
which are peaked at midwell.
Our model idealizes the electron density to be
midwell $\delta$~functions.}
\label{fig:tles}
\end{figure}

In experiments on TLESs, it is
extremely useful to be able to manipulate charges in the layers
with both a front (F) gate,
which we take to be closest to the left layer,
and a back (B) gate,
which we take to be closest to the right layer.
We parameterize these gate voltages in terms of neutralizing
charge densities $p_\alpha$ ($\alpha=$F,B), by defining 
\begin{equation}
V_{\alpha\rm{G}} = \frac{4\pi e}{\epsilon_0}
p_\alpha D_{\alpha\rm{G}} - V_{\alpha\rm{G}}^{(0)} ,
\label{eq:dp}
\end{equation}
where $\epsilon_0\approx 13$ is the dielectric constant for GaAs and 
$D_{\alpha\rm{G}}$ is the distance from the closest layer in the 
system to gate $\alpha$.
(For the systems fabricated by the Shayegan group,
$D_{\rm{FG}}\approx 0.45~\mu$m
and $D_{\rm{BG}}\approx 0.45~\rm{mm}$.\cite{shaypriv}) 
The last term represents an offset voltage.
This model for the triple-layer system is illustrated schematically
in Fig.~\ref{fig:tles}.

To test the appropriateness of such a model, and to determine
model parameters for the experimental system with which we 
will later compare our strong magnetic-field theory,
we have calculated
the dependence of the state of the triple-layer system 
on the front-gate voltage ($V_{\rm{FG}}$).
For this purpose, we have used 
a 2D version of the 3D local-density-functional (LDF)
theory.\cite{dftref}
The Kohn-Sham single-particle equations separate and yield three 
two-dimensional free-electron bands with minima at subband energies
that are determined
by solving the three-site discrete Schr\"odinger equation for the
$z$-direction degree of freedom.  In this equation, neighboring
layers are coupled by the tunneling matrix element $-t$ and the
LDF site energies are given (up to a common constant term) by 
\begin{eqnarray} 
\epsilon_1 &=&
\frac{2\pi e^2 d}{\epsilon_0} [(n_1-n_3)-(p_{\rm{F}}-p_{\rm{B}})] +
\mu_{\rm{xc}}(n_1) ,  \nonumber \\
\epsilon_2 &=& \epsilon_b +
\frac{2\pi e^2 d}{\epsilon_0} n_2 + \mu_{\rm{xc}}(n_2) , \nonumber \\
\epsilon_3 &=&
\frac{2\pi e^2 d}{\epsilon_0} [(n_3-n_1)-(p_{\rm{B}}-p_{\rm{F}})] +
\mu_{\rm{xc}}(n_3) .
\label{eq:siteeng}
\end{eqnarray} 
The ``bare'' middle-well on-site energy $\epsilon_b$ represents the 
difference in the size-quantization energy of the middle layer,
relative to that of the side layers.
It is negative (attractive) when the middle well is wider than
the side wells.
The Hartree terms are proportional to $2\pi e^2 d/\epsilon_0$.
The areal densities of the left, middle,
and right layers are denoted by
$n_1$, $n_2$, and $n_3$, respectively.
The exchange contribution to the exchange-correlation potential
$\mu_{\rm{xc}}$ of local-density-functional theory is
$\mu_{x}(n) = d[n\epsilon_{x}(n)]/dn = -e^2\sqrt{8n/\pi}/\epsilon_0$,
where $\epsilon_{x}(n)$ is the contribution to the exchange energy
per particle in a 2D electron gas
of uniform density $n$.\cite{dftref}
We have not included correlation-energy contributions to
$\mu_{\rm{xc}}$, but this could be added if desired.
The subband energies $E_\lambda$ are obtained
by diagonalizing the $3\times 3$ LDF Hamiltonian matrix.
The density in layer $j$ is given by
\begin{equation}
n_j = \sum_{\lambda=1}^3 N_\lambda |z^{(\lambda)}_j|^2 ,
\label{eq:density}
\end{equation} 
where
\begin{equation}
N_\lambda \equiv (E_{\rm{F}}-E_\lambda)
\nu_0 \Theta(E_{\rm{F}}-E_\lambda)
\label{eq:nlambda}
\end{equation}
is the areal-density contribution from the $\lambda$th subband,
$z^{(\lambda)}_j$ is the amplitude of the 
$\lambda$th subband wave function in layer $j$, and
$\nu_0 =m^{*} /\pi \hbar^2$ is the 2D electron-gas density of states
($m^{*}\approx 0.07m_{\rm{e}}$ is the band effective mass for GaAs).
Equations~(\ref{eq:siteeng}) and (\ref{eq:density}) have to 
be solved self-consistently, with the Fermi energy chosen so that
$\sum_\lambda N_\lambda = N_T = p_{\rm{F}} + p_{\rm{B}}$,
consistent with overall charge neutrality.

\section{Model Parameters}
\label{sec:prams}

Rough estimates of the size of the model parameters may be obtained
from simple arguments.
In the limit of infinitely strong barriers separating the layers,
the size-quantization energies are $\pi^2\hbar^2/2m^*w_j^2$,
where $w_j$ is the width of the $j$th quantum well.
For side-well widths of $15.4$~nm and a middle-well width of
$18.8$~nm,
the difference in the size-quantization energies gives
$\epsilon_b\sim -8$~meV for the TLES of Shayegan and co-workers 
that we study here.
We note that the next group of three electric subbands is higher
in energy by roughly
$3\pi^2\hbar^2/2m^*w^2\sim 50$~meV, and may be neglected in studying
the ground state and low-energy excitations of the TLES.
The tunneling energy may be crudely estimated from a
semiclassical (WKB) argument, by equating $2t/\hbar$
with the rate at which the wave-function amplitude
of an electron leaks out of its well by tunneling through
the confining barrier:
\begin{equation}
\frac{2t}{\hbar} \sim \frac{v}{2w} e^{-\kappa b}
= \frac{1}{\hbar} \frac{E_0}{\pi}  e^{-\kappa b},
\label{eq:twkb}
\end{equation}
where $v/2w$ is the frequency
with which an electron of average velocity
$v=\sqrt{2E_0/m^*}$ in a well of width $w\sim 17$~nm
hits the side of the well,
$E_0\approx\pi^2\hbar^2/2m^*w^2\sim 20$~meV is the energy of
the confined electron,
$\hbar\kappa=\sqrt{2m^*(V_0-E_0)}\approx\sqrt{2m^*V_0}$,
where\cite{shaypriv} $V_0\approx 1$~eV is the barrier height,
and $b\approx 1.3$~nm is the width of the barrier.
{}From this estimate,
the tunneling energy $t\sim 0.5$~meV, and is thus expected to be an
order of magnitude smaller than the on-site energy $\epsilon_b$.
We numerically solved the one-dimensional Schr\"odinger equation for
the triple quantum-well potential shown in Fig.~\ref{fig:tles} in an
effective-mass approximation to obtain the bound-state energies.
The three lowest energies determine $\epsilon_b$ and $t$ through
Eqs.~(\ref{eq:eps2}) and (\ref{eq:t}), and give
$\epsilon_b\approx -4.7$~meV and $t\approx 0.5$~meV.
We also find an energy separation of $34$~meV to the next group of
three electric subbands.

\begin{figure}[b]
\epsfxsize4.0in
\centerline{\epsffile{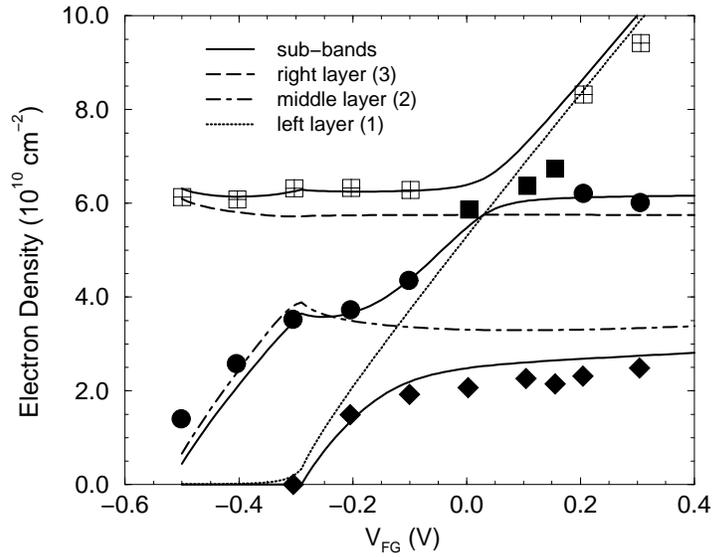}}
\caption{Shubnikov-de Haas data taken from Ref.~18,
and the subband (solid curves) and layer densities calculated from
the tight-binding model described in the text.}
\label{fig:sdh}
\end{figure}

Electrostatic energies tend to be larger than planar kinetic 
energies, tunneling energies, and exchange-correlation energies.
For the densities shown in Fig.~\ref{fig:sdh},
electrostatic energies are
typically about four times larger than planar kinetic energies,
about six times larger than the exchange energy,
and can be on the order of 100 times larger than the tunneling
energies, depending on the barrier widths.
As a result, Fig.~\ref{fig:sdh} can be understood qualitatively 
in an electrostatic approximation, where the electronic charge
resides entirely in the layer with the lowest
(Hartree plus ``bare'' on-site) site energy,
unless two or more layers are placed in equilibrium
by having the same energies.
For $4\pi e^2 d p_{\rm{F}}/\epsilon_0 \le \epsilon_b < 0$,
it follows that in this approximation, all electrons will occupy
the right layer, and $n_3=p_{\rm{B}}+p_{\rm{F}}$.
With increasing $p_{\rm{F}}$, charge will be added
first to the middle layer, until
$\epsilon_b + 4\pi e^2 d p_{\rm{F}}/\epsilon_0 \approx 0$.
As $p_{\rm{F}}$ is increased further,
all charge is added to the left layer.
These considerations provide two straightforward measurements
of $\epsilon_b$.
First, at the front-gate voltage $V_3\approx -0.3$~V
where the third subband first becomes occupied, electrostatic
considerations show that
\begin{equation}
\epsilon_b \approx -\frac{2\pi e^2 d}{\epsilon_0} N_2(V_3)
\approx -4.7\rm{~meV}
\label{eq:ebnl2}
\end{equation}
for $N_2(V_3)\approx 3.5\times 10^{10}\rm{~cm}^{-2}$
and $d=18.4$~nm.
Second, for fixed back-gate voltage,
the difference between $V_3$ and the value of the front-gate
voltage ($V_2\sim -0.6$~V) at which the second subband first
becomes occupied also measures $\epsilon_b$:
\begin{equation}
\epsilon_b \approx -\frac{1}{2} \frac{ed}{D_{\rm{FG}}} (V_3-V_2)
\approx -\frac{2\pi e^2 d}{\epsilon_0} (V_3 - V_2)
\frac{dN_T}{dV_{\rm{FG}}} \sim -6\rm{~meV}
\label{eq:ebfg}
\end{equation}
for $V_3-V_2 \sim 0.3$~V.
In Eq.~(\ref{eq:ebfg}), 
we have used the fact that $1/D_{\rm{FG}}$ is proportional
to the front-gate capacitance per unit area, $d(-eN_T)/dV_{\rm{FG}}$.

When the dependence of the chemical potential
on the density in each layer is taken into account,
there is a small correction to the electrostatic result.
This correction is measurable, and has been
exploited by Eisenstein and
co-workers\cite{eiscompress,jungwirth} in
the double-layer case to measure the compressibility of 
the electron-gas systems within each layer.  For example,\cite{jo}
features occur in the charge-density distribution when the 
density in one of the layers is very small, which reflect
the diverging (negative) compressibility in the low-density
limit of an electron gas.
In this picture, adding tunneling between the 
layers turns crossings of site energies 
into avoided crossings of subband energies, and smooths out cusps
in the dependences of the subband densities on the electric field. 

For double-layer systems, the hopping parameter $t$ 
is simply related to the subband energy separation when the gate
voltage is adjusted so that the two layers have equal density:
$ 2 t = E_2 -E_1 $.  This simple relationship is very helpful 
in characterizing experimental systems.  It is worth remarking 
that a similar simple relationship exists for triple-layer 
systems.  When $p_{\rm{F}}=p_{\rm{B}}$,
inversion symmetry guarantees that 
$n_1=n_3$ and $\epsilon_1=\epsilon_3$.  The Hamiltonian matrix 
can be readily diagonalized for this case with the result:
\begin{eqnarray}
E_1 &=& \frac{\epsilon_1 + \epsilon_2}{2}
- \sqrt{\left (\frac{\epsilon_1 -\epsilon_2}{2}\right )^2
+ 2 t^2}, \nonumber \\
E_2 &=& \epsilon_1, \nonumber \\
E_3 &=& \frac{\epsilon_1 + \epsilon_2}{2}
+ \sqrt{\left (\frac{\epsilon_1 -\epsilon_2}{2} \right)^2 + 2 t^2}.
\label{eqnarray:subeng}
\end{eqnarray}
The subband energies $E_\lambda$ can be determined up to
an overall constant by the sublayer occupancies $N_\lambda$,
which are obtained from the Shubnikov-de~Haas (SdH) experiments,
using Eq.~(\ref{eq:nlambda}).
Equation~(\ref{eqnarray:subeng}) can be solved to 
express $\epsilon_2 - \epsilon_1$ and $t$ in terms of the two 
independent subband energy differences.  We find that 
\begin{equation}
\epsilon_2 - \epsilon_1 = E_3 - 2 E_2 + E_1 ,
\label{eq:eps2}
\end{equation}
and that 
\begin{equation} 
t = \sqrt{\frac{(E_3-E_2)(E_2-E_1)}{2}}.
\label{eq:t}
\end{equation}
Since it is clear from Fig.~\ref{fig:sdh} that,
for the experimental system,
$E_3 - E_2 > E_2 - E_1$ when inversion symmetry is established,
we see immediately that 
$\epsilon_2 - \epsilon_1 > 0$ because of the electrostatic
energy cost of putting electrons in the middle layer. 
To determine $t$ experimentally, it is only necessary to identify
the gate voltage $V_{13}$ at which inversion symmetry 
is established, and use Eq.~(\ref{eq:t}).
$V_{13}$ may be determined in practice as
the point where ($N_1-N_2$)
is minimized, and electrostatic considerations give
$V_{13} \approx V_3 + 4\pi e D_{\rm{FG}} \bar{N}_2/\epsilon_0
\approx 0.1$~V,
where $\bar{N}_2\sim 6\times 10^{10}\rm{~cm}^{-2}$ is the
asymptotic value of $N_2$ at large $V_{\rm{FG}}$.
Unfortunately, the energy difference between $E_1$ and $E_2$
at the symmetric point is close to the
limit of resolution of the experiment.
An estimate of $t$ can be obtained from the minimum difference
between $N_2$ and $N_3$ (for $N_3>0$),
which occurs when the densities of layers 1 and 2 are equal.
This occurs at a front-gate voltage
$V_{12} \approx V_3 + 4\pi e D_{\rm{FG}} \bar{N}_3/\epsilon_0
\sim -0.1$~V,
where $\bar{N}_3\sim 3\times 10^{10}\rm{~cm}^{-2}$ is the
asymptotic value of $N_3$ at large $V_{\rm{FG}}$.
We obtain
\begin{equation}
t \approx \frac{1}{2} \rm{min}(E_3-E_2) =
\frac{\rm{min}(N_2-N_3)}{2\nu_0} \approx 0.45\rm{~meV}
\label{eq:t2}
\end{equation}
for $(N_2-N_3)\approx 2.5\times 10^{10}\rm{~cm}^{-2}$.

Subband densities can be extracted experimentally 
from weak-field SdH oscillation experiments.
The parameters of the model (the offset voltages,
the distances to the gates,
the size-quantization energy $\epsilon_b$,
and the hopping parameter $t$) may be determined by fitting 
to the gate-voltage dependence of the measured subband densities.
Experimental results are compared with a model fit
in Fig.~\ref{fig:sdh}.
The model calculations were performed with 
layer separation $d=18.4$~nm, the 
midwell to midwell distance in the experimental system.
The offset voltages $V_{\alpha\rm{G}}^{(0)}$ are consistent with
the condition\cite{lay2} that the electron density in the layers
is left-right (LR) symmetric for
$V_{\rm{FG}}=0.03$~V and $V_{\rm{BG}}=0$,
when the total electron density is $14.8\times 10^{10}\rm{~cm}^{-2}$.
Close agreement with experiment is obtained by choosing 
the tunneling and on-site energies to be within about $25\%$ of
$t=0.4{\rm ~meV}$ and $5\%$ of $\epsilon_b = -4.6{\rm ~meV}$.
The layer densities in Fig.~\ref{fig:sdh} are in close agreement with
the SdH data and with the layer densities obtained in
Ref.\onlinecite{lay} from 3D LDF calculations.
This level of agreement provides
us with the necessary confidence in our model, and determines with 
some assurance the 
model parameters for the system of immediate interest.
We note that this calculation (for $B_\perp=0$) explicitly 
neglects nonlocal interlayer exchange and correlation.
This approximation appears to be well justified at zero
magnetic field,\cite{leszek} but we will see that in the strong  
magnetic-field limit, nonlocal interlayer exchange is important.

\section{Quantum-Hall Ground State}
\label{sec:groundstate}
 
We now turn our attention to the strong magnetic-field limit.
Reference~\onlinecite{lay2} finds strong QHEs
at $\nu_{\rm{T}}=1$ and $2$; explaining the physics of the 
underlying incompressible states at these filling factors 
is the main objective of this paper.  The fact that a 
QHE occurs at these integer filling factors is,
at first sight, surprising.  To understand why, it is useful
to imagine repeating the calculations outlined in the 
previous section for $\nu_{\rm{T}}=1$.  The crucial difference
between the $B_\perp=0$ and strong-field situations in such
an independent-particle description is that,
because the kinetic energy is quenched,
the density of states consists of a $\delta$~function 
at the subband energies in the strong-field case, whereas
it is a constant above the subband energy in the $B_\perp=0$ 
case.  As a result, the distribution of density between
the three layers at $\nu=1$,
which we have emphasized is dictated 
largely by electrostatics, can be expressed simply in terms 
of the subband wave function:
\begin{equation}
n_{j} = (2 \pi \ell^2)^{-1} |z^{(1)}_j|^2.
\label{eq:sfdens}
\end{equation}
In order to have the charge distributed relatively equally
among all the layers, the 
site energies cannot differ by more than $\sim t$.  The 
difference between subband energies, which would give the 
QH activation gap in such a theory, would then 
also be $\sim t$.  A gap of this size might be reduced 
or possibly eliminated by disorder in the samples.
In the experiments, however, the observed gaps can be much
larger than $t$.   It seems clear that the explanation for  
these QHEs must lie in the interaction 
physics of the triple-layer system at strong fields, and that 
the gap would exist even if $t$ were zero.    
Further evidence is found in the experimental observation that 
as the ratio of side-layer to middle-layer electron density
is increased, the $\nu_{\rm{T}}=1$ state collapses,
but the $\nu_{\rm{T}}=2$ state state becomes stronger.
We shall show that these observations are consistent
with the behavior
expected if spontaneous interlayer phase coherence occurs in 
these triple-layer systems.  In the following, we focus 
on the case $\nu_{\rm{T}}=1$;
the $\nu_{\rm{T}}=2$ case is simply related by
a particle-hole transformation.\cite{fractoo}
We shall also assume that the system has three layers,
although most aspects generalize in an obvious way to systems
with more than three layers.

The single Slater-determinant states considered in this section 
have the form
\begin{equation}
|\Psi\rangle = \prod^{N_\phi}_X \left ( \sum_{j=1}^{3} z_j
{\hat c}^\dagger_{jX} \right ) |0\rangle ,
\label{eq:psi3}
\end{equation}
where $z_j$ is the component in layer $j$ of a normalized
three-subband wave function, and ${\hat c}^\dagger_{jX}$ 
is the second-quantization operator that 
creates an electron in layer $j$, in the lowest Landau-level
Landau-gauge state with guiding-center coordinate $X$.
This wave function is a full Landau-level state for the 
subband with state vector $Z^\dagger = (z_1^*,z_2^*,z_3^*)$.  
In our HFA, we allow $Z$ to be varied to minimize the energy.
This variation generally results in a broken-symmetry
ground state, since $z_i^* z_j \ne 0$ even when $t=0$;
i.e., the HF ground state has 
spontaneous interlayer phase coherence.
For double-layer systems, the broken symmetry is robust
under appropriate circumstances,
and exists in the exact quantum-mechanical ground state.
We expect spontaneous interlayer phase coherence to be similarly 
robust for triple-layer systems.

It is convenient to define the density matrix
\begin{equation}
\rho_{jk}(X) = \langle \Psi| \hat{c}^\dagger_{jX}
\hat{c}_{kX} |\Psi\rangle =  z^*_j z_k ,
\label{eq:defrho}
\end{equation}
where $| \Psi \rangle $ is the coherent
ground-state wave function, Eq.~(\ref{eq:psi3}).
Note that the filling factor of layer $j$ is $\nu_j=\rho_{jj}$.
In terms of the density matrix, the HF total energy is
\begin{eqnarray}
E_{\rm{HF}} & = & \sum_{jX} \biggl\{
- \sum_k t_{jk} \rho_{jk}(X)
+ \epsilon_j \rho_{jj}(X)
- \sum_{\alpha Y} \bar{\nu}_\alpha D_{\alpha j}(Y) \rho_{jj}(X)
\nonumber\\
& + &  \frac{1}{2} \sum_{kY} \left[
D_{jk}(X-Y) \rho_{kk}(Y) \rho_{jj}(X)
-  E_{jk}(X-Y) \rho_{kj}(Y) \rho_{jk}(X) \right]
\biggr\} ,
\label{eqnarray:ehf}
\end{eqnarray}
where $t_{jk}$ denotes the tunneling energy
between layers $j$ and $k$,
$\epsilon_j$ is the on-site size-quantization energy
of layer $j$, and $\alpha$ indexes
neutralizing planes of charges with areal charge density
$ep_\alpha = e\bar{\nu}_\alpha/2\pi \ell^2$
produced by remote ionized donors or gates.
The unit of length is the magnetic length,
$\ell=\sqrt{\hbar c/eB}$.
As an aside, we note that in the long-wavelength limit,
$E_{\rm{HF}}$ has the form of a $CP_{N-1}$ model\cite{rajaraman}
when expressed in terms of the $z_j$.\cite{cpn}

In the lowest Landau level,
the Coulomb interaction between electrons in layers $j$ and $k$
enters through the direct term $D_{jk}(X-Y)$ and
the exchange term \mbox{$E_{jk}(X-Y)$.}
These quantities are conveniently expressed in terms of
projected Fourier transforms: e.g.,
\begin{equation}
D_{jk}(X) = \int \frac{d^2q}{(2\pi)^2}
D_{jk}(q) e^{\frac{1}{4}q^2 \ell^2}
\int d^2r \langle X|{\bf r}\rangle
e^{i{\bf q}\cdot{\bf r}} \langle {\bf r}|X\rangle ,
\label{eq:dxy}
\end{equation}
where
\begin{equation}
D_{jk}(q) =
\frac{2\pi e^2}{q} e^{-qd|j-k|}  e^{-\frac{1}{2}q^2 \ell^2} ,
\label{eq:dq}
\end{equation}
and
\begin{equation}
E_{jk}(q) = \frac{\ell^2}{2\pi} \int d^2p D_{jk}(p)
e^{i({\hat {\bf z}}\cdot{\bf p}\times{\bf q}) \ell^2} .
\label{eq:eq}
\end{equation}
The layers are located at $z=dj$,
where $j$ is in general a real number, unless
all the layers are separated by integer multiples of $d$.
We have neglected the finite thickness of the electron
wave functions in the $z$ direction,
although this effect could be included, if desired.

When $\nu_{\rm{T}}=1$, it follows from Eq.~(\ref{eq:psi3}) that
\begin{equation}
\rho_{jk}(X) \rho_{kj}(X) = \rho_{jj}(X) \rho_{kk}(X) .
\label{eq:rhojkkj}
\end{equation}
This relation, which follows from the phase-coherent nature of
the assumed ground state, is of great practical importance to us,
because unlike in the $B_\perp=0$ case,
it allows us to easily express the exchange contribution to the
HF ground-state energy $E_{\rm{HF}}^0$ in terms of
the layer occupancies:
\begin{equation}
\frac{E_{\rm{HF}}^0}{N_\phi} = \sum_{jk} \left [
-t_{jk} \sqrt{\nu_j\nu_k}
+ \delta_{jk} \left ( \epsilon_j -
  \sum_\alpha \bar{\nu}_\alpha \bar{D}_{\alpha j} \right ) \nu_j
+ \frac{1}{2} (\bar{D}_{jk}-\bar{E}_{jk}) \nu_j\nu_k \right ] ,
\label{eq:ehfn}
\end{equation}
where
\begin{equation}
{\bar D}_{jk} = -\left[ D_0(q=0) - D_{|j-k|}(q=0)
\right]/2\pi \ell^2 = -v_c \frac{d}{\ell} |j-k|
\label{eq:djk}
\end{equation}
with
$v_c\equiv e^2/\epsilon_0\ell$,
and
\begin{equation}
{\bar E}_{|j-k|} \equiv \sum_X E_{jk}(X) =
\frac{E_{jk}(q=0)}{2\pi \ell^2} =
v_c \int_0^\infty dx e^{-\frac{1}{2}x^2} e^{-x|j-k|d/\ell} .
\label{eq:ejk}
\end{equation}
Equation~(\ref{eq:rhojkkj}) also allows us to write
\begin{equation}
\rho_{jk}(X) = \sqrt{\nu_j\nu_k} e^{i\phi_{jk}(X)} .
\label{eq:rhojjkk}
\end{equation}
In the ground state, $\rho_{jk}$ is independent of $X$,
so that we may take $e^{i\phi_j(X)}=1$. 
Since $\sum_j\nu_j=1$, the HF ground state is 
in general determined by two parameters. 
We will concentrate below on the situation where the triple-layer
system is symmetric, i.e., $\bar{\nu}_{\rm{F}}=\bar{\nu}_{\rm{B}}$.
Then the system has inversion symmetry around the middle well,
$\nu_1=\nu_3$, and the ground state is completely fixed by $\nu_2$.

The layer occupancies can be calculated
by minimizing Eq.~(\ref{eq:ehfn});
equivalently, the following simple argument may be used
in the absence of tunneling.
For $\epsilon_b$ sufficiently large, all electrons reside in
the side layers, closest to the gates.
Suppose that both side layers (1 and 3) are occupied, and
imagine moving an electron from layer 1 to the middle layer (2).
The energy gained by moving to the middle has three components:
the on-site energy $\epsilon_b$, the Hartree energy
$\bar{D_1}[(\nu_1+\nu_3)-(\bar{\nu}_{\rm{F}}+\bar{\nu}_{\rm{B}})]$
due to the direct Coulomb interaction with the side layers and gates,
and the exchange energy $-\bar{E_1}(\nu_1+\nu_3)$
of the middle electron with the side layers.
The energy lost by moving to the middle also has three contributions:
the Hartree energy $\bar{D_2}(\nu_3-\bar{\nu}_{\rm{B}})$
of a layer-1 electron
with the electrons in layer 3 and the gates,
the on-site exchange energy $-\bar{E_0}\nu_1$ in layer 1,
and the exchange energy $-\bar{E_2}\nu_3$ of an electron in layer 1
with those in layer 3.
Specializing to the case of $\nu_{\rm{T}}=1$,
and using $\bar{D_2}=2\bar{D_1}$,
the energy cost to move an electron to the middle layer is seen to be
\begin{equation}
\Delta E = \epsilon_b +
\bar{D_1}[(\nu_1-\nu_3)-(\bar{\nu}_{\rm{F}}-\bar{\nu}_{\rm{B}})]
           + (\bar{E_0}-\bar{E_1})\nu_1
	   - (\bar{E_1}-\bar{E_2})\nu_2 .
\label{eq:delemax}
\end{equation}
The middle layer will be occupied when $\Delta E < 0$.
For the case of equal side-layer densities,
this happens when $\epsilon_b < \epsilon_b^{\rm{max}}$, where
\begin{equation}
\epsilon_b^{\rm{max}} = -\frac{1}{2}
(\bar{E_0}-2\bar{E_1}+\bar{E_2}) .
\label{eq:ebmax}
\end{equation}
A similar argument can be used to find the value
$\epsilon_b^{\rm{min}}$ of the on-site energy below which
all electrons reside in the middle layer:
\begin{equation}
\epsilon_b^{\rm{min}} = -(-\bar{D_1}-\bar{E_0}+\bar{E_1}) .
\label{eq:ebmin}
\end{equation}

Equations~(\ref{eq:ebmax}) and (\ref{eq:ebmin}) show that
near-neighbor interlayer exchange
($\bar{E_1}$) plays an essential role 
in increasing the size of the interval where all three layers 
are occupied and spontaneous triple-layer phase coherence occurs.
The local-density approximation for exchange, commonly 
used in electronic structure calculations, fails qualitatively
for triple-layer (and double-layer) systems in the QH regime,
because it does not include the effects of interlayer exchange,
included here through $\bar{E_j}$ for $j>0$.
The on-site exchange ($\bar{E_0}$) favors maximizing the
charge of individual layers, and $\bar{E_2}$ favors
next-nearest-neighbor occupancy.
For $\epsilon_b^{\rm{min}} \le \epsilon_b \le \epsilon_b^{\rm{max}}$,
the middle-well occupancy decreases linearly with $\epsilon_b$,
so that
\begin{equation}
\nu_2 = \left [ \frac{\epsilon_b^{\rm{max}} - \epsilon_b}
       {\epsilon_b^{\rm{max}} - \epsilon_b^{\rm{min}}} \right ] .
\label{eq:nu2}
\end{equation}
When $\nu_2=1$, all the electrons are in the middle well,
and we have, in effect, a single-layer system.  When
$\nu_2=0$, all the electrons are shared between the outside layers,
and we have a double-layer system.  For $ 0 < \nu_2 < 1$,
charge exists in all three layers and the HFA ground state has 
triple-layer coherence. 

We may use the HF ground-state calculation of $\nu_2$ in
Eq.~(\ref{eq:nu2}) to define a phase diagram in the space
of model parameters in the absence of interlayer tunneling.
For symmetric triple-layer systems,
the state of the system is determined by the 
middle-well size-quantization energy $\epsilon_b$
in units of the Coulomb energy $v_c\sim 10$~meV,
and by the interlayer spacing $d$ in units of $\ell$.
We consider the limit $t/v_c \ll 1$, which according to our
analysis of the SdH data is satisfied by the device studied
in Ref.\onlinecite{lay2}, and set $t=0$.
The region of stability of the triple-layer coherent state 
at $t=0$ is bounded by the dotted and dashed (upper and lower)
lines of Fig.~\ref{fig:phase_diags}(a).
These two lines are defined by the equations 
$\nu_2= 1$ and $\nu_2 = 0$.
At fixed $d$, the system transforms with increasing $|\epsilon_b|$,
first from a double-layer system to a triple-layer system,
and finally to 
a single-layer system with all the charge in the middle layer.
At fixed $\epsilon_b$, these transformations occur in the 
opposite order with increasing $d$, as electrostatic 
considerations become more dominant.   
Figure~\ref{fig:phase_diags}(b)
shows the same plot for $\nu_{\rm{T}}=2$,
obtained from a calculation nearly identical
to that of the $\nu_{\rm{T}}=1$ case.
If, in a given sample, front and back gate voltages 
are adjusted simultaneously to maintain LR symmetry as
the total density changes, the system will follow
a line in this phase diagram that moves downward and toward 
the right with increasing magnetic field,
as illustrated in Fig.~\ref{fig:phase_diags}.
By this procedure, the triple-layer regime should be acessible
to experimental study in typical TLESs.
The HF ground-state calculation by itself suggests that the entire
region between the dotted and dashed lines in
Fig.~\ref{fig:phase_diags} might support a triple-layer
coherent state.
However, as we show in the following section,
a stability analysis using the
TDHF equations shows that for sufficiently large interlayer spacing,
the phase-coherent state cannot be the ground state.

In closing this section, we remark that the distribution of
electrons between the three layers can be markedly different
for the same gate voltages and sample parameters
in the QH regime, as compared to the $B_\perp=0$ case.
Reference\onlinecite{lay2} found, using their 3D LDF
calculations, that for their triple-layer sample at symmetry,
with a total density of $14.8\times 10^{10} \rm{~cm}^{-2}$,
the ratio of the density of electrons in the middle layer to
the total density was
$\nu_2/\nu_{\rm{T}}\approx 0.22$ when $B_\perp=0$.
The 2D tight-binding model of the previous section gives
the same result.   For the same model parameters, we find that 
in the $\nu_{\rm{T}}=1$ phase-coherent triple-layer QH (3LQH) state,
$\nu_2/\nu_{\rm{T}}\approx 0.27$,
while for $\nu_{\rm{T}}=2$ we obtain
$\nu_2/\nu_{\rm{T}}\approx 0.09$.
The low value of $\nu_2/\nu_{\rm{T}}$
for $\nu_{\rm{T}}=2$ is due to the fact
that for inversion-symmetric triple-layer systems,
the $\lambda=2$ subband wave function has
no weight in the middle layer.
Thus $B_\perp$ can have a large effect on the ratio of
layer densities.

\section{Time-dependent Hartree-Fock collective modes}
\label{sec:colmodes}

In the TDHF description of the collective behavior, the variational
wave function for the ground and excited states of the system
both have the HF form, Eq.~(\ref{eq:psi3}),
but the density matrix elements $\rho_{jk}$
are allowed to have spatial and time dependence.
In particular, the low-lying collective modes arise from slow 
long-wavelength variations of the phase differences $\phi_{jk}(X)$.
In practice,
there are several ways to implement the TDHF approximation:
diagrammatically, by path-integral methods, or by various equation of
motion (EOM) approaches.
TDHF collective modes for a triple-layer system
were calculated previously by Fertig\cite{fertigcm} using 
a diagrammatic approach, for the unphysical case of equidistant,
individually charge-neutral layers with periodic boundary conditions,
with the tunneling and Coulomb energies between end
layers taken to be the same as between neighboring layers.
Here we briefly describe a calculation based on the EOM of the 
density matrix.  For the sake of generality, we use the language of 
an $N$-layer system, although we will apply our results to the case
$N=3$.

Following Ref.\onlinecite{cote}, we define
the projected density-matrix operator
\begin{equation}
\hat{\rho}_{jk}({\bf q}) = e^{\frac{1}{4}q^2 \ell^2}
\sum_{XY} \hat{c}^\dagger_{jX}
\left [ \int d^2r \langle X|{\bf r}\rangle
e^{-i{\bf q}\cdot{\bf r}} \langle {\bf r}|Y\rangle
\right ] \hat{c}_{kY} ,
\label{eq:rhoq}
\end{equation}
and the density-density response function
\begin{equation}
\chi_{jklm}({\bf q},t) = \frac{1}{i}
\langle T \hat{\rho}_{jk}({\bf q},t)
\hat{\rho}_{jk}(-{\bf q},0) \rangle ,
\label{eq:denden}
\end{equation}
where $T$ denotes the time-ordering operator.
The physical quantities of interest to us are a function of
$q=|{\bf q}|$, due to the isotropy of the system.
The HFA to $\chi$ is obtained from the EOM
\begin{equation}
i\hbar\partial_t \hat{\rho}_{jk}({\bf q})
 = [ \hat{\rho}_{jk}({\bf q}) , \hat{H}_{\rm{HF}} ] ,
\label{eq:hfeom}
\end{equation}
where $\hat{H}_{\rm{HF}}$ is the HF Hamiltonian.
The commutator is evaluated by use of the identity
\begin{equation}
[\hat{\rho}_{jk}({\bf p}) , \hat{\rho}_{lm}({\bf q})]
= \delta_{kl} \hat{\rho}_{jm}({\bf p}+{\bf q})
e^{i({\hat {\bf z}}\cdot{\bf p}\times{\bf q}) \ell^2}
- \delta_{mj} \hat{\rho}_{lk}({\bf q}+{\bf p})
e^{i({\hat {\bf z}}\cdot{\bf q}\times{\bf p}) \ell^2} .
\label{eq:commut}
\end{equation}

In the TDHFA, $\chi(q,\omega)$ is obtained from the
HF $\chi^{(0)}$ by including the effects of the
Coulomb interaction between a particle in layer $j$ and
a hole in layer $k$ with total in-plane momentum $q$,
within a generalized random-phase approximation.
This results in a Dyson equation for $\chi$
in terms of the direct and exchange interaction between
a particle and hole.  In matrix notation,
\begin{equation}
\chi = \chi^{(0)} + \chi^{(0)} \frac{1}{\hbar} W \chi ,
\label{eq:dysoneq}
\end{equation}
where $W=H-X$ is the sum of the Hartree and Fock contributions to
the particle-hole interaction, given by
\begin{equation}
H_{jklm}(q) = \delta_{jk} \delta_{lm}
\frac{D_{jl}(q)}{2\pi \ell^2} ,
\label{eq:hart}
\end{equation}
and
\begin{equation}
X_{jklm}(q) = \delta_{jm} \delta_{kl}
\frac{E_{jl}(q)}{2\pi \ell^2} .
\label{eq:fock}
\end{equation}
Since the layer indices each have $N$ possible values,
Eq.~(\ref{eq:dysoneq}) may be solved numerically by inverting an
$N^2\times N^2$ matrix.

\begin{figure}[t]
\epsfxsize4.0in
\centerline{\epsffile{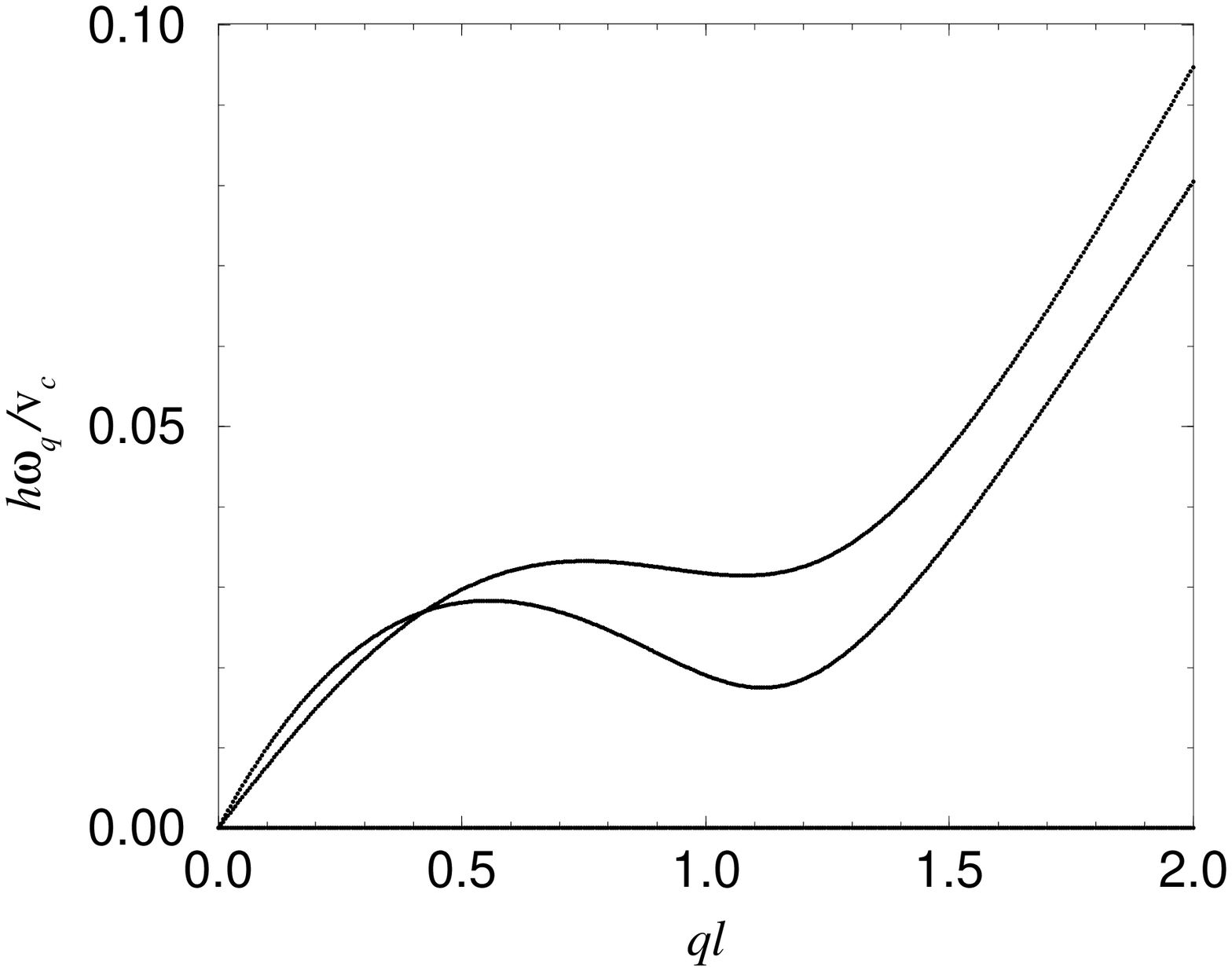}}
\caption{Collective modes of the $\nu_{\rm{T}}=1$ 3LQH coherent state
for ($t=0$,$-\epsilon_b/v_c=0.75$, $d/\ell=1.6$),
which has $\nu_2=0.838$.
The lack of an avoided crossing at $q\ell\sim 0.4$ is due to the 
assumed left-right inversion symmetry of the system.}
\label{fig:coll_modes}
\end{figure}

The collective modes are obtained from the poles of
$\chi(q,\omega)$ that have nonzero residues.
The results of a sample calculation for
($t=0$,$-\epsilon_b/v_c=0.75$, $d/\ell=1.6$),
which has $\nu_2=0.838$,
are shown in Fig.~\ref{fig:coll_modes}.
(The collective mode for $\nu_{\rm{T}}=2$
can be obtained from that for
$\nu_{\rm{T}}=1$ by particle-hole conjugation, using
$\nu_2\rightarrow 1-\nu_2$.) 
As seen from Fig.~\ref{fig:coll_modes}, there are
two collective modes, which correspond to linear superpositions
of variations in $\phi_{12}(X)$ and $\phi_{23}(X)$.
As the spacing $d$ between the layers increases,
one of the collective modes softens and eventually becomes
unstable at a wavevector $q \sim \ell^{-1}$. 
This signals the onset of a charge-density wave instability
in the HFA, on which we comment further below.
The lack of an anticrossing repulsion at the point
(away from $q=0$) where the two collective-mode frequencies
are equal is a special feature of the assumed LR
symmetry of the charge distribution in the TLES.
When this symmetry is present, the collective modes are excitations
from a subband state ($\lambda=1$) that has even parity
to subband states that have even ($\lambda=2$)
or odd ($\lambda=3$) parity.
The opposite parity between the two final states provides a 
selection rule that prevents the two collective modes from mixing.

So far, we have focused on the case where the tunneling $t_{jk}$
is negligible compared to the Coulomb interaction energy scale
$v_c$, and can be set to zero.
This results in gapless Goldstone modes, as shown, for example,
in Fig.~\ref{fig:coll_modes}.
(When $1<\nu_{\rm{T}}<N-1$, gapped collective modes exist,
even in the absence of tunneling.)
The tunneling energy can be varied by changing the thickness of
the barrier between the quantum wells.
The $[U(1)]^{N-1}$ invariance associated with the freedom
to choose the ${N-1}$ relative phases $\phi_{j,j+1}$
in each layer is broken once the electronic states in different
layers are coupled by tunneling.
The loss of this invariance gives a gap of order $2t$
(for nearest-neighbor $t_{jk}=t$) in the collective-mode spectrum
as $q\rightarrow 0$.

\begin{figure}[b]
\epsfxsize4.0in
\centerline{\epsffile{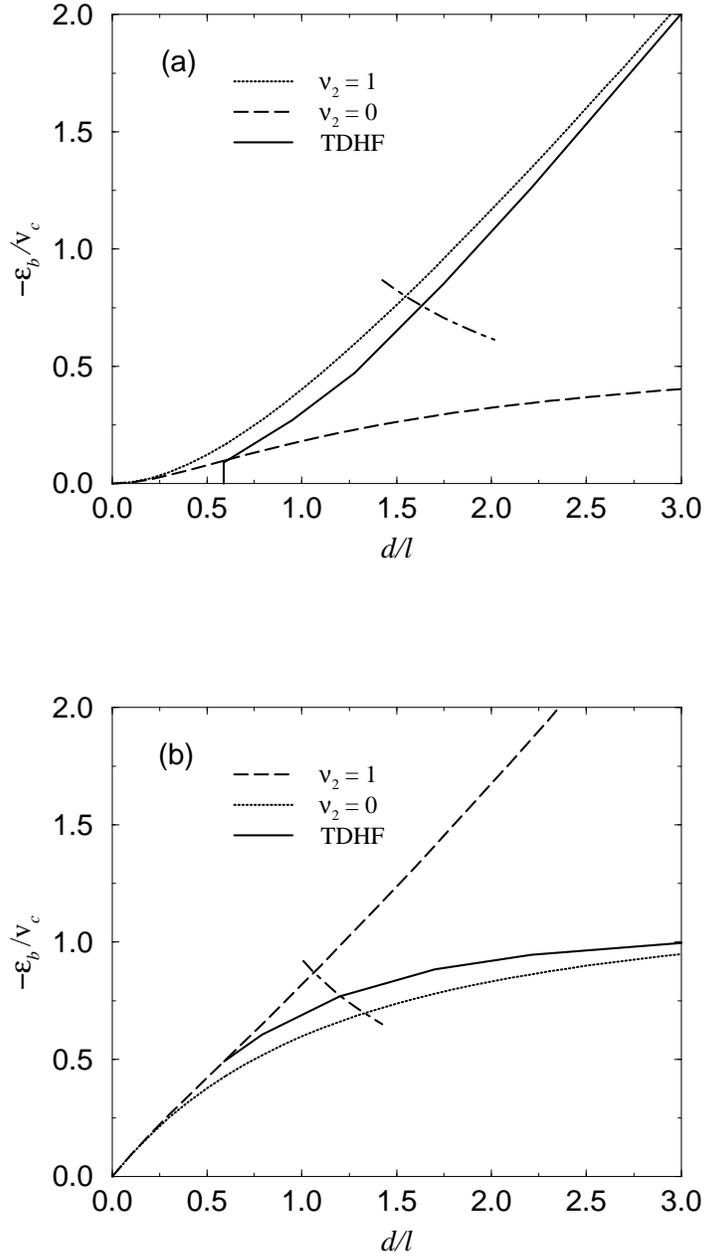}}
\caption{(a) Phase diagram for the symmetric $\nu_{\rm{T}}=1$
3LQH coherent state.
         (b) Phase diagram for the symmetric $\nu_{\rm{T}}=2$
3LQH coherent state.
The mean-field regions of stability lie between the dotted and
solid lines.
All calculations are for $t=0$, for the case of equal left and
right layer densities.
The dot-dashed lines represent values of $(d/\ell,-\epsilon_b/v_c)$
obtained for a sample with $d=18$~nm, for total density in the
range of $1$ to $2\times 10^{11}\rm{~cm}^{-2}$, using
$\epsilon_b=-8$~meV for $\nu_{\rm{T}}=1$, and
$\epsilon_b=-6$~meV for $\nu_{\rm{T}}=2$.}
\label{fig:phase_diags}
\end{figure}

In our TDHF collective-mode calculations, we find that for 
fixed $\epsilon_b$, the collective-mode spectrum in the 
triple-layer regime softens with increasing $d$, and that,
except for sufficiently small values of $d/\ell$,
an instability occurs before the $\nu_2=0$ line in 
Fig.~\ref{fig:phase_diags}(a) is reached.
The first collective mode that goes soft corresponds to an
excitation from the filled $\lambda=1$ subband to the empty
$\lambda=2$ subband, which has all the charge on the outside
layers; this is the favored arrangement for the charge at large $d$.
This instability appears in the TDHFA
as an imaginary-valued collective-mode frequency. 
Based on extensive calculations performed previously for
the case of double-layer systems,\cite{cote} the broken translational
symmetry HF ground states that are found for larger values
of $d$ will quickly lose their interlayer phase coherence, and 
the charge gap (incompressibility) necessary for the integer
QHE will rapidly go to zero.
In our view, the broken translational symmetry of the 
state on the large-$d$ side of this instability is likely
to be an artifact of the HFA, which can 
enhance intralayer correlations only by breaking translational
symmetry.  When quantum fluctuations are included, the broken
translational symmetry of this state is likely to be lost,
but in our view the loss of interlayer phase coherence and the 
vanishing of the QH charge gap will remain.
We therefore use the location of the TDHF instabilities 
as an estimate of the layer separation at which spontaneous
coherence and the integer QHE are lost.
The results are shown in Fig.~\ref{fig:phase_diags}(a)
for $\nu_{\rm{T}}=1$.
The solid line shows the border of the TDHF instability.
For $\nu_{\rm{T}}=1$,
the region between the upper dotted line ($\nu_2$=1)
where the middle layer is fully occupied, and the solid line, where
the TDHF instability occurs, is the resulting estimate of 
the region in parameter space where a 3LQH coherent state occurs.
This shows that as $d/\ell$ increases,
the $\nu_{\rm{T}}=1$ phase-coherent
3LQH state is more likely to be stable when the middle layer has
{\it more} electrons than the side layers.\cite{dcnote}
Figure~\ref{fig:phase_diags}(b) shows the phase diagram
for $\nu_{\rm{T}}=2$.
In this case, the 3LQH state at larger $d/\ell$ is likely to be
more stable when the middle layer has {\it fewer} electrons
than the side layers.
These behaviors are seen in the experiments of Ref.\onlinecite{lay2}.
The dot-dashed lines in Fig.~\ref{fig:phase_diags} represent
values of $(d/\ell,-\epsilon_b/v_c)$ for a hypothetical sample
with $d=18$~nm,
as the total density is varied from
$1$ to $2\times 10^{11}\rm{~cm}^{-2}$,
where we have used $\epsilon_b=-8$~meV for $\nu_{\rm{T}}=1$ and
$\epsilon_b=-6$~meV for $\nu_{\rm{T}}=2$.
As the total density and $B_\perp$ are increased while
keeping the total filling factor $\nu_{\rm{T}}$ constant,
the dot-dashed
lines in Fig.~\ref{fig:phase_diags} show that the ratio of
middle-layer density to side-layer density decreases.  
In all cases, the presence of interlayer tunneling
will enlarge the region where a triple-layer QHE 
is expected to occur. 
We emphasize that the TDHF instability with increasing $d$ is
driven by the increasing relative importance of intralayer 
correlations compared to interlayer correlations.
At large enough layer separations, the interlayer coherence that
provides for good interlayer correlations will be lost, along with
the charge gap necessary for the QHE.
This scenario for the disappearance of the QHE with increasing layer
separation does not require that the TLES make a transition
to a bilayer system, as was hypothesized in Ref.~\onlinecite{lay2}.

\section{Concluding Remarks}
\label{sec:summary}

We have shown that TLESs,
such as those fabricated by Shayegan and co-workers,
can be described using a simple tight-binding model, which 
allows them to be characterized in terms of a small number
of parameters.  The tight-binding model 
is able to quantitatively account for weak-field SdH 
experimental results for the dependence of the three subband 
energies on the gate voltage.  Using this model, we have 
estimated the dependence of the ground state of the system on
model parameters for filling factors $\nu_{\rm{T}}=1$ and $2$.
We propose that in triple-layer systems, as in double-layer
systems, the QHE can occur at integer 
total filling factors, even when the charge is
distributed among all three layers and there is no hopping
between the layers.
(Tunneling would be required for noninteracting electrons
to produce a gap at integer filling factors.)
These QHEs occur because of the formation of broken-symmetry
ground states with spontaneous interlayer coherence.

The distribution of charge between the three layers of the system
is typically determined predominantly by electrostatic
considerations, and states can occur with the electrons
distributed among one, two, or three layers.
Our HF theory of the ground state shows that for physically
accessible regions of the middle-well
on-site energy and layer separation,
a triple-layer phase coherent state is possible.
The stability of this state was estimated using the TDHFA,
and a phase diagram was constructed,
delimiting the regions in the parameter space
of the TLES for which triple-layer coherence is likely to
be found.  The QH states studied by Shayegan and co-workers
at $\nu_{\rm{T}}=1$ and $2$ show the behavior
expected for phase-coherent states.
In particular, they exhibit the QHE, even for small
tunneling energies and unequal
layer densities.  Additionally, the $\nu_{\rm{T}}=1$ QHE 
is suppressed by increasing occupancy of the side wells at
the expense of the middle well,
while for $\nu_{\rm{T}}=2$, the opposite is 
true, in agreement with our findings.  Future detailed comparisons
between our phase diagram and experiment will be 
facilitated by the possibility of following lines in
the phase diagram for a single sample, by adjusting gate
voltages so that inversion-symmetric $\nu_{\rm{T}}=1$ and 
$2$ states occur for a range of magnetic fields.  In 
any such quantitative comparison, it will be necessary to 
approximately account for the finite widths of the 
individual quantum wells, which we have not done here. 

As in the case of the double-layer coherent state,
a useful footprint of interlayer coherence is 
unusual sensitivity to small tilts of the magnetic field
away from the normal to the layers.\cite{murphy}
The most convincing evidence for triple-layer coherence would be
the observation of a strong suppression in the activation energy of
the 3LQH states due to the application of a moderate
parallel magnetic field.  The state of the triple-layer 
system in a tilted magnetic field is most informatively 
described using a field-theoretical approach.
We have calculated the magnitude of the parallel field required
to suppress the activation energy as a function of the layer
density, for both triple-layer and unbalanced double-layer systems.
The effect of a parallel field on triple-layer states,
charged order-parameter textures of triple-layer states, and 
other related properties of triple-layer phase-coherent states will
be discussed in a forthcoming paper.\cite{cpn}

\section{acknowledgments}
\label{sec:ack}

It is a pleasure to acknowledge useful conversations with
Steve Girvin, Mansour Shayegan, and Sanjeev Shukla.
C.B.H. thanks Charles~A.~Hanna
for inspiration and invaluable assistance.
This work was supported by the National Science Foundation 
under Grant No. DMR94-16906.

\newpage

\addcontentsline{toc}{part}{Figure Captions}

\newpage

\begin{figure}[t]
\epsfxsize4.0in
\centerline{\epsffile{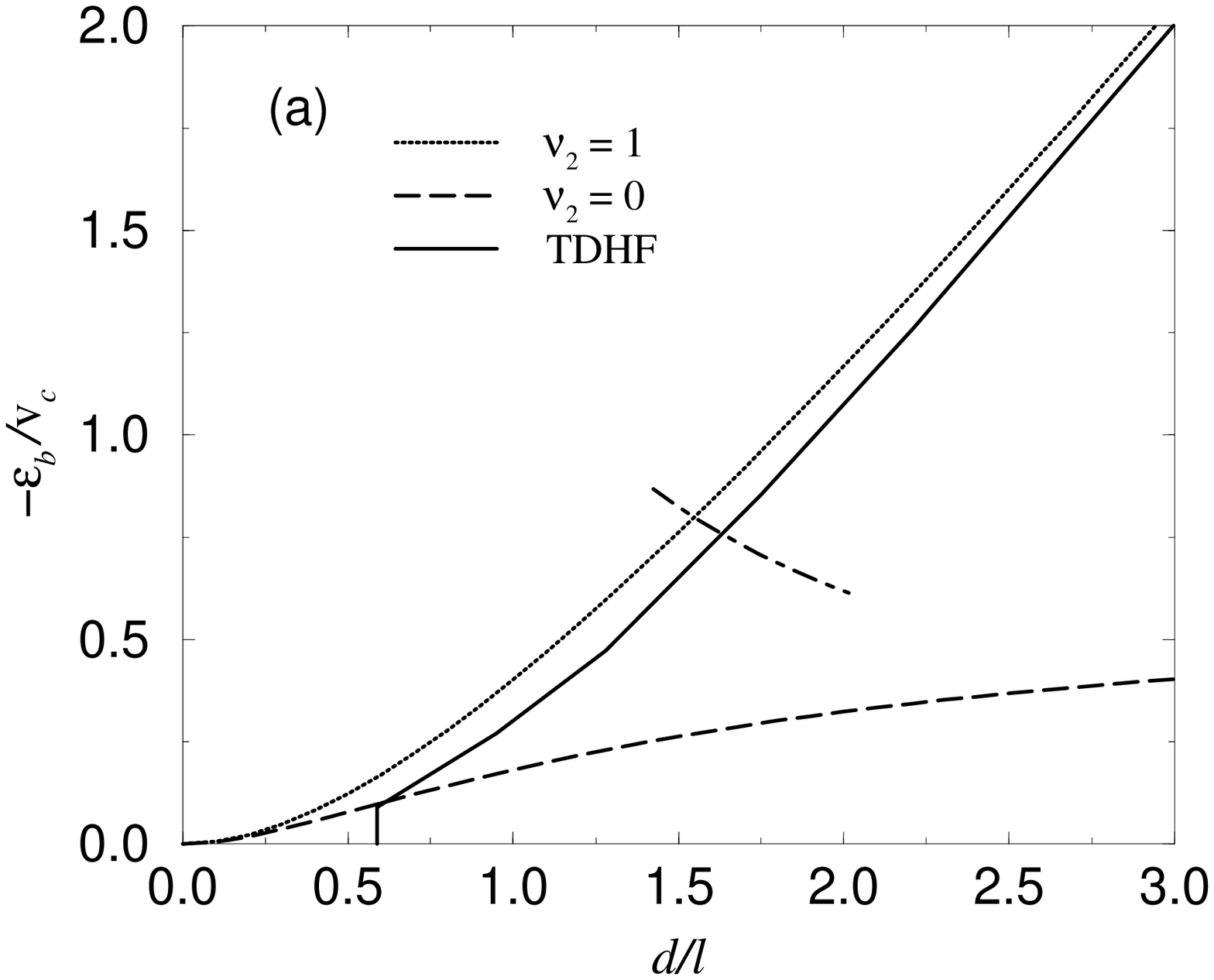}}
\end{figure}

\newpage

\begin{references}
\bibitem{fractoo} Spontaneous phase coherence can 
occur at nonintegral filling factors as well, but we restrict
our attention here to the integer case where the mobility 
requirements for experimental systems are not as severe.
This formalism also describes interlayer phase coherence in an
$N$-layer sytem at filling factors $\nu_{\rm{T}}=1$ and $N-1$.
\end{references}
\end{document}